\documentclass[12 pt]{article}
\usepackage{cite,amsmath,amsfonts,amsthm,fullpage}
\usepackage{youngtab}
\usepackage{cite,amsmath,amsfonts,amsthm,fullpage, amssymb, mathtools, url}
\usepackage[usenames,svgnames]{xcolor}
\usepackage{youngtab, enumitem, comment, mathdots, graphicx, float}
\usepackage[symbol]{footmisc}
\usepackage[pagebackref=true]{hyperref} 

\theoremstyle{plain}
\theoremstyle{definition}
\newtheorem{theorem}{Theorem}[section]
\newtheorem{lemma}[theorem]{Lemma}
\newtheorem{definition-theorem}[theorem]{Definition-Theorem}
\newtheorem{definition-proposition}[theorem]{Definition-Proposition}

\newtheorem{corollary}[theorem]{Corollary}
\newtheorem{example}{Example}[section]
\newtheorem{examples}{Example}[subsection]

\newtheorem{remark}{Remark}[section]
\newtheorem{remarks}{Remarks}[section]

\numberwithin{equation}{section} 

\def\tr{\mathrm {tr}}
\def\det{\mathrm {det}}

\def\&{&{\hskip -20pt}}

\def\be{\begin{equation}}
\def\ee{\end{equation}}

\def\bea{\begin{eqnarray}}
\def\eea{\end{eqnarray}}

\def\bt{\begin{theorem}}
\def\et{\end{theorem}}

\def\bex{\begin{example}\small \rm}
\def\eex{\end{example}}

\def\bexs{\begin{examples}\small \rm}
\def\eexs{\end{examples}}

\def\br{\begin{remark}\small \rm}
\def\er{\end{remark}}

\def\pb{\mathbf{p}}

\def\tilp{\tilde{p}}
\def\tilpb{\tilde{\bf{p}}}
\def\tN{\textsc{N}}
\def\tv{\textsc{v}}
\def\tf{\textsc{f}}
\def\te{\textsc{e}}
\def\tZ{\textsc{Z}}
\def\tI{\textsc{I}}

\def\bp{\begin{Proposition}\rm}
\def\ep{\end{Proposition}}
\def\bc{\begin{corollary}}
\def\ec{\end{corollary}}
\def\bl{\begin{lemma}\em}
\def\el{\end{lemma}}
\def\be{\begin{equation}}
\def\ee{\end{equation}}
\def\br{\begin{remark}\rm\small}
\def\er{\end{remark}}
\def\brs{\begin{remarks}.\\ \rm\
\begin{enumerate}}
\def\ers{\end{enumerate}\end{remarks}}
\def\bea{\begin{eqnarray}}
\def\eea{\end{eqnarray}}

\begin{document}

\begin{center}
\begin{Large}\fontfamily{cmss}
\fontsize{17pt}{27pt}
\selectfont
	\textbf{Integrals of tau functions I: Round dance tau function and multi-matrix integrals}
	\end{Large}
	
\bigskip \bigskip
\begin{large} 
A. Yu. Orlov$^{1,2}$\footnote[2]{e-mail:orlovs55@mail.ru} 
 \end{large}
 \\
\bigskip

\begin{small}
$^{1}${\em Shirshov Institute of Oceanology, Russian Academy of Science, Nahimovskii Prospekt 36, Moscow 117997, Russia }\\
$^{2}${\em 
A.I. Alikhanov  Institute for Theoretical and Experimental Physics 
of NRC Kurchatov Institute, B. Cheremushkinskaya, 25, Moscow, 117259, Russia}
\end{small}
 \end{center}

\medskip

\begin{abstract}
The simplest nontrivial tau functions of the Toda lattice and the $\tN$-component Toda lattice are compared
in their applications to multimatrix integrals.
\end{abstract}
\bigskip


\section{Introduction}

I want to dedicate this article to Andrei Pogrebkov in connection with his 75th birthday. Andrei has made 
notable and varied contributions to the development of the theory of integrable systems - the vertex operator 
(the work by Pogrebkov and Sushko 1974, late published \cite{PogrebkovSushko}), correctly defined Hamiltonian formalism, the 
inverse problem of scattering theory, a detailed study of a number of specific integrable equations, and a 
number of other often very original approaches to integrability theory.
The text below represents a slightly different direction,
but I am grateful for the always inspiring discussions with Andrey on this topic.

\section{The simplest nontrivial tau functions of the multicomponent TL}

The well-known Cauchy-Littlewood relation
can be written as
\be\label{CLittle}
e^{-\sum_{m>0}\frac 1m p_m \tilp_m}=1+\sum_{\kappa > 0}
\sum_{\alpha_1>\cdots >\alpha_\kappa\atop \beta_1>\cdots >\beta_\kappa}
s_{(\alpha|\beta)}(\pb) s_{(\beta|\alpha)}(\tilpb) = \tau_o(\pb,\tilpb)
\ee
Here $\pb=(p_1,p_2,\dots)$ and $\tilde{\pb}=(\tilde{p}_1,\tilde{p}_2,\dots)$ are two infinite sets of
variables and $s_{(\alpha|\beta)}$ denotes the Schur polynomial \cite{Mac} in the power sum variables labeled
by a partition $(\alpha|\beta)$ written in the Frobenius coordinates $\alpha=(\alpha_1,\dots,\alpha_\kappa)$ 
and $\beta=(\beta_1,\dots,\beta_\kappa)$.
This is also the simplest example of the Toda lattice (TL) \cite{Mikhailov} tau functions \cite{JM},\cite{UT} 
where the variables $\pb$ and $\tilpb$ play the role of the so-called higher times. 
\footnote{For higher times, 
other designations are sometimes used, for example, $n x_n$ instead of $p_n$, the sign can be also different. 
To fix the correspondence we write the Appendix.} It is written in 
form of series in the Schur functions as it was  worked out in \cite{Takasaki-Schur}, \cite{TI}.
The straightforward generalization of this expression to the $\tN$-component Toda lattice ($\tN$-TL) yields
\be\label{round_dance}
\tau_o(\pb^{1},\tilpb^{1},\dots,\pb^{\tN},\tilpb^{\tN})=
1+ \sum_{\kappa > 0} 
\sum_{\alpha^1_1>\cdots >\alpha^1_\kappa,\atop \beta^1_1>\cdots >\beta^1_\kappa}
\cdots
\sum_{\alpha^\tN_1>\cdots >\alpha^\tN_\kappa,\atop \beta^\tN_1>\cdots >\beta^\tN_\kappa}
 \,\prod_{1\le i\le \tN}^{\circlearrowleft}
 s_{(\alpha^{i}|\beta^{i})}(\pb^{i})s_{(\beta^{i}|\alpha^{i+1})}(\tilpb^{i})
\ee
where each set of the higher times is denoted by $\pb^i$ and $\tilpb^i$, $i=1,\dots,\tN$.
The symbol $\circlearrowleft$ above the symbol of the product means that we suppose the 
condition $\alpha^{\tN+1}=\alpha^1$, 
which closes the chain of Schur functions in a circle which explains the motivation of the name ``round dance''.
I like this expression and it somehow reminds me of Matisse's painting ``Dance''.
There is a beautiful expression for series (\ref{round_dance}), 
similar to the left side in (\ref{CLittle}), see formula (38) in \cite{MirMorHook}.

\section{Graphs and matrix models \label{Graphs}}

Consider the set of complex $GL_N$-matrices $Z_1,\dots,Z_n$ and their Hermitian conjugates
$Z_{-1},\dots,Z_{-n}$, where $Z_{-i}:=Z_i^\dag$. We multiply each $Z_i$ by $C_i\in GL_N$ on the right,
$Z_i\to (Z_iC_i)$, where $i=\pm 1,\dots,\pm n$.
The matrices $Z_i$ will be called random, and the matrices $C_i$ will be called source ones.
The elements of random matrices are distributed according to the Gauss law as follows:
\be\label{corfun}
\langle (Z_a)_{ij}(Z_{a'})_{i'j'}\rangle =\frac1N \delta_{a+a',0}\delta_{ii'}\delta_{jj'}
\ee
Such correlation functions are obtained for the so-called Ginibre complex multi-matrix ensembles
\cite{Mehta},\cite{Ipsen},\cite{Strahov}.
To calculate various correlation functions, we use Wick's rule. We will study special correlation functions, 
which are obtained as follows:

We denote the collection of random matrices by $\tZ$.
We consider the set of $GL_N$ matrices
$$
W_1(\tZ),\dots,W_\tv(\tZ)
$$
where each $W_a(\tZ)$ is the product of $Z_iC_i$ pairs. Each product $W_a(\tZ)$ is defined up to a cyclic permutation.
We set the condition: each pair enters one and
only once to one of the members of the set 
$\{W_a(\tZ),a=1,\dots,\tv\}$. In this case, there is 
a two-dimensional orientable surface $\Omega$ and an embedded graph\footnote{For embedded
graphs, there are other names: "thick graphs" and "ribbon graphs" in physical literature and 
"maps" in mathematical literature} with $n$ numbered edges and with $\tf$ numbered vertices, drawn 
on $\Omega$ as follows. Each edge labeled $|i|$ ($|i|=1,\dots,n$) consists of two half-edges with 
numbers $i$ and $-i$ and associate the half-edge $i$ with the matrix $Z_i$. Each vertex labeled $a$ 
is connected to $W_a(Z)$ as follows: if we move clockwise, then the outgoing half-edges will have the 
same numbers as the random matrices that are included in the product $W_a(\tZ)$, if we read it from left 
to right.
Then we number the corners of the faces of the graph as follows:
when traversing a vertex clockwise, each half-edge with the number $i$ ($i=\pm 1,\dots,\pm n$) 
is followed by an angle with the same number. Then in corner $i$ we place $C_i$.
We will use the notation $W_a(\tI)$ which denote $W_a(\tZ)$ where we put all matrices $Z_i$ to 
be the identity matrix.
We see that $W_a(\tI)$ can be called the monodromy of the vertex $a$ and $W_a(\tZ)$ the monodromy dressed by random 
matrices. 
Suppose that the Euler characteristic of $\Omega$ is $\te$, $\te=\tf-n+\tv$, here $\tf$ is the number of 
faces of $\Gamma$. 
Now we will number all the faces of $\Gamma$. We also assign a monodromy to each face 
which is a matrix defined 
up to the cyclic order. Monodromy $W_b^*(\tI)$ of the face with number $b$ is introduced as the product of all source 
matrices $C_i$ at the corners of the face in the order in which we encounter them when going around the face 
counterclockwise. 
The dressed monodromy of the face $W_a(\tZ)$ is obtained from $W_a(\tI)$, when we replace each matrix $C_i$ that 
enters $W_a(I)$ by $Z_iC_i$.

It can be seen that the dressed monodromies of faces of $\Gamma$ is the dressed monodromy of the vertices of 
the dual graph $\Gamma^*$.
 This construction was analyzed in the works \cite{NO2020},\cite{NO2020F},\cite{2DYM},\cite{AOV}.

 \paragraph{Integrals of products of the Schur functions}
 For any sets of partitions $\{\lambda^a,a=1,\dots,\tv\}$ and $\{\lambda^b,b=1,\dots,\tv$\}
we have
\be\label{evV}
\langle \prod_{a=1}^\tv s_{\lambda^a}(W_a(\tZ) \rangle = c
\delta_{\lambda}\prod_{b=1}^\tf s_\lambda(W_b^*(\tI))
\ee
\be\label{evF}
\langle \prod_{b=1}^\tf s_{\lambda^b}(W^*_b(\tZ)) \rangle = c 
\delta_{\lambda}\prod_{a=1}^\tv s_\lambda(W_a(\tI))
\ee
\be
c=\left(\frac{\dim\lambda}{|\lambda|!}\right)^{-n} N^{-n|\lambda|}
\ee
where $\dim\lambda$ is the dimension of the irreducable representation of the symmetric group 
$S_d$ ($d=|\lambda|!$) labeled by $\lambda$. The symbol $\delta_\lambda$ is equal to $1$ if all 
partitions are equal to each other: $\lambda^1=\lambda^2=\cdots =:\lambda$,
  otherwise $\delta_\lambda =0$. Notice the diagonalization property for the products 
  of the Schur function when we evaluate the expectation $\langle \rangle$.
\br
These nice relations can be compared with the relations for embedded graphs \cite{ZL}
\be\label{on_base1}
\sigma \alpha = \varphi
\ee
\be\label{on_base2}
\sigma \varphi = \alpha
\ee
where,  $\sigma,\alpha,\varphi\in S_{2n}$ and $\sigma$ is the involution without fix points, $\alpha$ 
is the product of cycles related 
to vertices and $\varphi$ is the product of cycles

\er

For a given matrix $X$ and a partition $\mu=(\mu_1,\dots,\mu_{\ell})$ we introduce the notation
\be\label{powerSum}
\pb_\mu(X) = \tr(X)^{\mu_1}\cdots \tr(X)^{\mu_\ell}
\ee
One can associate $\pb_\mu(X)$ with a set of polygons consisting of $\mu_1$-gons,...,$\mu_\ell$-gons.
Suppose that an $i$-gon is included in this set $m_i$ times.
Introduce number $\texttt{Aut}(\mu):=\prod_{i>0} m_i! i^{m_i}$, which can be called the number of automorphisms
of this collection: number of permutations
polygons with the same number of edges multiplied by the number of rotations of each polygon.

One can derive (\ref{evV})-(\ref{evF}) in different ways, in particular \cite{NO2020}, starting from the relation
\be\label{polygons}
\langle \prod_{b=1}^\tf \frac{\pb_{\mu^b}(W^*_b(\tZ))}{|\texttt{Aut}(\mu^b)|} \rangle = 
\sum_{\nu^1,\dots,\nu^\tv} {\cal{H}}_\Omega \left(\mu^1,\dots,\mu^\tf,\nu^1,\dots,\nu^\tv \right)
\prod_{a=1}^\tv \pb_{\nu^a}(W_a(\tI)) 
\ee
where all partitions $\mu^1,\dots,\mu^\tf,\nu^1,\dots,\nu^\tv $ are of the same weight, say $d$, $d\le N$. 
Here ${\cal{H}}_\Omega$ is the Hurwitz number, $\Omega$ the base surface and partitions  
$\mu^1,\dots,\mu^\tf,\nu^1,\dots,\nu^\tv $ are the ramification profiles. 

The meaning of the formula (\ref{polygons}) is as follows.
For all partitions $\mu^1,\nu^1,\cdots $ equal to the partition $(1)$, this formula describes the base surface, 
glued from polygons associated with $\{\tr(W^*_b(\tZ)),b=1,\dots,\tf\}$, see (\ref{on_base2}). 
Expression for $\pb_\mu$ corresponds to the set of polygons on the covering surface,
which correspond to the set of profiles $\mu^b$ in the center (in the "capital") of the polygon $b$.
Wick's rule gives all possible gluings of the collection of polygons given by $\pb_\nu$. Each way of gluing leads 
to a set $\{\nu^a,a=1,\dots,\tv \}$ of branching profiles at the vertices. The number of ways to glue 
polygons (up to the automorphisms) for given sets of all ramification profiles 
$\mu^1,\dots,\mu^\tf,\nu^1,\dots,\nu^\tv$ is the Hurwitz number in its geometric meaning.

Then from the Frobenius formula for the Hurwitz numbers and from the orthogonality relations of the characters
symmetric group, we get (\ref{evF}) and (\ref{evV}). Another way to get
(\ref{evF}) and (\ref{evV}) - consistent use of well-known formulas for integrating Schur functions
(using analytic continuation of parameters), see for instance \cite{2DYM}.

If instead of the Ginibra ensemble we consider ensembles of unitary matrices (circular ensembles), 
then the relations (\ref{evV}) and (\ref{evF}) will change slightly, in particular, the dimension of the representation 
of the symmetric group will have to be replaced by the dimension of the representation of the linear group.
On the related topics see also \cite{O-2004-New} and \cite{Alexandrov}.

Actually we need only formula (\ref{evV}) for the next part.

\section{Round dance tay function and multimatrix Ginibre ensemble}

Now consider the construction of the previous section and in formala (\ref{round_dance}) 
(where we replace $\tN$ by $\tv$ to avoid a mess in the usage of the capital N)  we choose the times $
\tilpb^a$ as follows 
\be\label{p-W}
\tilp^{(a)}_k=\tr(W_a(\tZ))^k,\quad a=1,\dots,\tv
\ee
We get
\be\label{generating}
\langle \tau(\pb^1,\tilpb^1),\dots , \pb^\tv,\tilpb^\tv) \rangle=
\sum_\lambda \left(\frac{\dim\lambda}{|\lambda|!}\right)^{-n} N^{-n|\lambda|}
\prod_{a=1}^\tv s_\lambda(\pb^a)
\prod_{b=1}^\tf s_\lambda(W_b^*(\tI))
\ee
The series (\ref{generating}) appeared in the works \cite{MM1},\cite{MM5},\cite{MM4},\cite{MM3}, 
\cite{AMMN-2011},\cite{AMMN-2014} as a generating function for Hurwitz numbers. To achieve these
series we chose the condition (\ref{p-W}).

Note that, apparently, the right-hand side can only be a tau function of the hypergeometric type of the 
TL hierarchy \cite{KMMM},\cite{Kharchev98},\cite{OS-2000},\cite{OS-TMP}. It's not an exact statement, 
but I don't see any other possibilities at the moment. 
That is, there may be a tau function of the form 
\be
\sum_\lambda s_\lambda(\pb^1)s_\lambda(\pb^2)\prod_{(i,j)\in\lambda} r(j-i)
\ee
with some function $r$.

What are our chances of getting it. First, the Euler characteristic of the base surface $\Omega$ must 
be equal to two, therefore we have a graph $\Gamma$ on a Riemann sphere. 
Then only two sets of free parameters should remain. Let's look at all the possibilities, there are three of them:

(1) two sets among $\pb^a$ are free, all others have a form
\be\label{p(a)}
\pb^i=\pb(a_i):=(a_i,a_i,\dots),\quad i=1,\dots,k
\ee
(in the considered case $k\le\tv$)
or coinside with
\be\label{p-infty}
\pb^i=\pb_\infty:=(1,0,0,\dots),\quad i=1,\dots,\tv-k
\ee
where $a$ is a complex number. 
In this case the spectrum of matrices $W^*_b$, $b=1,\dots,\tf$ should be
\be\label{spectrum}
\texttt{Spectr}\left[W_b(\tI)\right]=(\underbrace{1,\dots,1}_{N_b},\underbrace{0,\dots,0}_{N-N_b})
\ee
with some $N_b$. In this case the right hand side of (\ref{generating}) is as follows 
\be\label{hyp1}
 \sum_\lambda N^{-n|\lambda|}s_\lambda(\pb^1)s_\lambda(\pb^2)\prod_{(i,j)\in\lambda}
 \prod_{c=1}^k (a_c+j-i)\prod_{b=1}^\tv (N_b+j-i)
\ee
About specializations (\ref{p(a)}),(\ref{p-infty}) see \cite{OS-2000}), \cite{O-2004-New},
and see \cite{2DYM},\cite{AOV} on (\ref{spectrum}).

Among these tau functions, one can be found, which was studied in \cite{ChekhovAmbjorn}.
The authors use rectangular matrices $Z$, in our approach this is the choice of source matrices
which may degenerated. The graph $\Gamma^*$ in this case is just an open chain.

Tau functions of type (\ref{hyp1}) for the first time were used in \cite{Goulden-Jackson-2008},
then in \cite{Uspehi-KazarianLando}
in the context of combinatorial problems and also
in \cite{PaqHarnad}, \cite{HO-2014},\cite{NOBKP} as the generating functions of Hurwitz numbers of certain type.

(2) We have one free set, say, $\pb^1$, all others have forms (\ref{p-infty}) and (\ref{p(a)}).
We have a single free monodromy, say, $W^*_1$, all others have their spectrum as in (\ref{spectrum}).
In this case the right hand side of (\ref{generating}) is 
\be\label{hyp2}
 \sum_\lambda N^{-n|\lambda|}s_\lambda(\pb^1)s_\lambda(\pb^2(W^*_1(\tI)))\prod_{(i,j)\in\lambda}
 \prod_{c=1}^{k+1} (a_c+j-i)\prod_{b=2}^\tv (N_b+j-i)
\ee
It is written as a function of the set of the higher times $\pb^1$ and as a function in the so-called Miwa
variables given by the spectrum of $W^*_1(\tI)$

(3) We two free monodromies, say, these are $W^*_1(\tI))$ and $W^*_2(\tI))$ while
the spectrum of all others are given by (\ref{spectrum}). Each $\pb^a$ is either (\ref{p(a)}) or 
(\ref{p-infty}). In this case we have the right hand side of (\ref{generating}) written in Miwa
variables as
\be\label{hyp3}
 \sum_\lambda N^{-n|\lambda|}s_\lambda(\pb^1(W^*_1(\tI)))s_\lambda(\pb^2(W^*_2(\tI)))\prod_{(i,j)\in\lambda}
 \prod_{c=1}^{k+2} (a_c+j-i)\prod_{b=3}^\tv (N_b+j-i)
\ee

\bigskip
\bigskip
\noindent 
\small{ {\it Acknowledgements.} 
I would like to thank Andrei D. Mironov for the inspiring discussion.
The work was supported by the Russian Science
Foundation (Grant No.20-12-00195).
\bigskip



\appendix
\section{Partitions and Schur functions}

The partition is a set of integers $\lambda=(\lambda_1,\dots,\lambda_\ell)$, where
$\lambda_1 \ge \cdots \lambda_\ell > 0$ and $\lambda_i$ are called parts of $\lambda$.
The sum of the parts is called the weight of $\lambda$ and is usually denoted as $|\lambda|$.
  There is a well-known notion of Young diagrams $\lambda$ whose row lengths are equal to parts of $\lambda$.
We accept the agreement as in \cite{Mac}.
Let us denote the horizontal lengths of the part of the strings that starts at the $ii$ node (and does 
not include it) and goes to the right as $\alpha_i$, and the vertical part of the column that starts at 
(not including) the same node and goes down we denote $\beta_i$. In Frobenius coordinates, $\lambda$ is written 
as $(\alpha|\beta)=(\alpha_1,\dots,\alpha_k|\beta_1,\dots,\beta_k)$, where $k$ is the number of nodes on 
the main diagonal of the Young diagram, see \cite{Mac} for details.

The following relations define Schur polynomials (Schur functions) as a function of
power sum variables $\pb=(p_1,p_2,\dots)$:
$$
e^{\sum_{k>0}\frac 1k p_k x^k}=\sum_{k\ge 0} s_{(k)}(\pb)
$$
$$
s_{\lambda}(\pb)=\det\left[s_{(\lambda_i-i+j)}(\pb) \right]_{i,j}
$$
$$
{\rm In\,case}\quad p_m=\tr(X)^m \quad {\rm one\, can\, write}\quad 
s_\lambda(\pb(X))=:s_\lambda(X)
$$
We have
$$
s_\lambda(\pb)=\frac{\dim\lambda}{|\lambda|!}\sum_{\mu} \varphi_\lambda(\mu) \pb_\mu,
\quad \dim\lambda =\frac{\prod_{i<j}(\lambda_i-\lambda_j-i+j)}{\prod_i (\lambda_i-i+\ell)!}
$$
The known orthogonality relations for the factors $\varphi_\lambda$ (which are the normalized characters of the 
symmetric group in representation $\lambda$) \cite{Mac} are as follows:
\be\label{orth1}
 \zeta_{\Delta} \sum_{\lambda} \left(\frac{{\rm\dim}\lambda}{d!}\right)^2 
\varphi_\lambda(\mu)\varphi_\lambda(\Delta) =
  \delta_{\Delta,\mu} 
\ee
and
\be\label{orth2}
\left(\frac{{\rm\dim}\lambda}{d!}\right)^2
\sum_{\Delta} \zeta_\Delta\varphi_\lambda(\Delta)\varphi_\mu(\Delta) =
\delta_{\lambda,\mu}
\ee
where $d=|\Delta|=|\lambda|$ and if each part $i$ enters $m_i$ times the partition $\Delta$ we define
\be\label{z_Delta}
\zeta_\Delta=\prod_{i}m_i!i^{m_i}
\ee
(The same number was denoted $|\texttt{Aut}(\Delta)|$ in the paragraph 
"Integrals of products of the Schur functions").
All details can be found in \cite{Mac}.  

The Frobenius formula for Hurwitz numbers mentions in Section \ref{Graphs} is
$$
{\cal{H}}_\Omega(\Delta^1,\dots,\Delta^m)=\sum_\lambda\left(\frac{\dim\lambda}{|\lambda|!}\right)^{\textsc{e}}
\varphi_\lambda(\Delta^1)\cdots \varphi_\lambda(\Delta^m)
$$
where $\textsc{e}$ is the Euler characteristic of the base surface $\Omega$, the
partitions $\Delta^1,\dots,\Delta^m$ is the set of the ramification profiles in branch points, see 
the textbook \cite{ZL} for the details.

\section{Fermions. Multicomponent tau function}

I will only briefly give the notation and some formulas. Fermion operators $\psi^{(a)}_n,\psi^{\dag(a)}_n$ 
where $a=1,\dots,\tN $, $n\in\mathbb{Z}$,
satisfy the relations
$$
[\psi^{(a)}_n,\psi^{(b)}_m]_+=0=[\psi^{(a)\dag}_n,\psi^{(b\dag)}_m]_+,
\quad [\psi^{(a)\dag}_n,\psi^{(b)}_m]_+ =\delta_{a,b}\delta_{n,m}
$$
and 
$$
\langle 0|\psi^{(a)}_{n}  =\langle 0|\psi^{\dag(a)}_{-n-1} = 
0 =\psi^{(a)}_{-n-1}|0\rangle=\psi^{\dag(a)}_{n}|0\rangle, \quad n>0
$$

The wonderful observation of Kyoto school (see for instance \cite{JM}) is the following relation
\be
s_{(\alpha|\beta)}(\pb^{a})= (-1)^{\sum_{i=1}^k \beta_i}\langle 0|e^{\sum_{k>0} \frac1k p^{(a)}_k J_k^{(a)}}
\sum_{i=1}^k \psi_{\alpha_i}^{(a)}\psi^{\dag(a)}_{-\beta_i-1}  |0\rangle,\quad 
\ee
which is also known as bosonization formula (a version of this formula). Here 
$$J_n^{(a)}=\sum_{i\in\mathbb{Z}} \psi^{(a)}_i\psi^{\dag(a)}_{n+i}$$. We also need the property 
$s_{(\alpha|\beta)}(\pb)=(-1)^{|(\alpha|\beta))|}s_{(\beta|\alpha)}(-\pb)$.

Following \cite{JM} we introduce $2\tN$-component KP ($\tN$-component Toda) tau function in form of a 
fermionic expectation value which we choose as follows: 
\be
\langle 0|e^{\sum_{a}\sum_{k>0}(-1)^{a+1}\frac 1k p_k J^{(a)}_k} 
e^{\sum_{1\le a\le 2\tN}^\circlearrowleft \sum_i \psi^{\dag(a)}_{-i-1}\psi^{(a+1)}_{i}} |0\rangle
=:\tau_o(\pb^1,\dots,\pb^\tN)
\ee
where $\circlearrowleft $ means that $\psi^{(2\tN+1)}_{-i-1}:=\psi^{(1)}_{-i-1}$.
We call it round dance tau function. Expanding the exponential in the last expression in  Taylor series and using 
the relations above, we get (\ref{round_dance}). In case $\tN=1$ we get the right hand side of (\ref{CLittle}).

\section{Complex Ginibre ensemble}

On this subject there is an extensive literature, for instance see lists of references in 
\cite{Strahov,Ipsen,Alexandrov,ChekhovAmbjorn,NO2020,NO2020F,AOV}.

Let us consider integrals over $N\times N$ complex matrices $Z_1,\dots,Z_n$ where the measure is defined as
\be\label{CGEns-measure}
d\Omega(Z_1,\dots,Z_n)= c_N^n
\prod_{i=1}^n\prod_{a,b=1}^N d\Re (Z_i)_{ab}d\Im (Z_i)_{ab}\text{e}^{-N|(Z_i)_{ab}|^2}
\ee
where the integration domain is $\mathbb{C}^{N^2}\times \cdots \times\mathbb{C}^{N^2}$ and where $c_N^n$
is the normalization
constant defined via $\int d \Omega(Z_1,\dots,Z_n)=1$. We get (\ref{corfun}) for the correlation functions of the 
entries of the matrices $Z_1,\dots,Z_n$.

The set of $n$ $N\times N$ complex matrices
 with measure (\ref{CGEns-measure}) is called the set of {\it $n$ independent complex
Ginibre ensembles}. Such ensembles have wide applications in physics  and in
information transfer theory. 

\end{document}